\input{aipcheck}

\documentclass[
    ,final            
  ]
  {aipproc}

\layoutstyle{6x9}

\begin{document}

\title{New results on SIDIS SSA from Jefferson Lab}


\author{H.~Avakian, P.~Bosted, V.~Burkert, L.~Elouadrhiri {\small for the} CLAS \small Collaboration}{
  address={Jefferson Lab, Newport News, VA 23606, USA}
}

\begin{abstract}
We present studies of single-spin  and double-spin  asymmetries in 
semi-inclusive electroproduction of pions using the CEBAF 6
GeV polarized electron beam.
Kinematic dependences of single and double spin asymmetries
have been measured in a wide kinematic range at CLAS with a 
polarized  NH$_3$ target. 
Significant target-spin $\sin2\phi$ and $\sin\phi$ asymmetries 
have been observed.
The hypothesis of factorization has been
tested with $z$-dependence of
the double spin asymmetry.

\end{abstract}

\maketitle


Single-Spin Asymmetries (SSAs) in azimuthal distributions of final state
particles in semi-inclusive deep inelastic scattering  play a 
crucial role in the study of transverse momentum distributions
of quarks in the nucleon and provide access to the
orbital angular momentum of quarks.
Large SSAs, observed for decades in hadronic reactions  have been among 
the most difficult phenomena to understand from first principles in QCD.
Recently, significant SSAs were reported in semi-inclusive DIS (SIDIS) 
by the HERMES collaboration at HERA \cite{HER,HERtrans} 
for longitudinally and transversely polarized targets, 
and by the CLAS collaboration  
with a polarized beam \cite{classsa}.

Two fundamental mechanisms have been identified leading to SSAs in hard processes,
the Sivers mechanism \cite{SIV,AM,BRODY,COLL,JIY}, which generates 
an asymmetry in the distribution of quarks due to orbital motion
of partons, and the Collins
mechanism \cite{COL,TM}, which generates 
an asymmetry  during the hadronization of quarks.
\smallskip

The HERMES Collaboration has recently measured a transverse spin
asymmetry in SIDIS providing the cleanest evidence to date
for the existence of a non-zero Collins function \cite{HERtrans}, 
which describes the fragmentation of a transversely polarized quark into pions.
This finding is supported by the preliminary data from BELLE \cite{BELLE} indicating
a non-zero Collins effect.
The large target SSA in semi-inclusive pion production measured at 
CLAS and analyzed in terms of the Collins 
fragmentation \cite{efrem2004}, also indicate a significant Collins function.

For a longitudinally polarized target the only azimuthal asymmetry
arising in leading order is the $\sin2\phi$ moment \cite{KO,TM,KM96}, 
involving the transverse momentum
dependent (TMD) Collins fragmentation function $H_1^\perp$ \cite{COL} and 
the Mulders distribution function $h_{1L}^\perp$ \cite{RS,TM}, describing
 the transverse polarization of quarks in a longitudinally polarized proton \cite{KO,TM,KM96}. 
The same distribution function is accessible in double polarized
Drell-Yan, where it gives rise to a $\cos2\phi$ azimuthal moment 
in the cross section \cite{TM0}.

Single and double spin asymmetries in SIDIS have been measured using
the  CLAS \cite{CLAS}
in Hall B at Jefferson Lab, a 5.7 GeV longitudinally polarized 
electron beam, a longitudinally polarized proton (NH$_3$) target.
The average beam polarization was  $0.73 \pm 0.03$ and
the average target polarization was  $0.72 \pm 0.05$. 
The open acceptance of CLAS and a single electron trigger ensured
event recording for a large sample of SIDIS $\pi^+$, $\pi^0$,
and $\pi^-$ events. 

{\vskip -0.5cm
\begin{figure}[h]
\hspace{-20mm}
 \begin{minipage}[b]{6.0cm}
  \includegraphics[height=.22\textheight,width=3.0in]{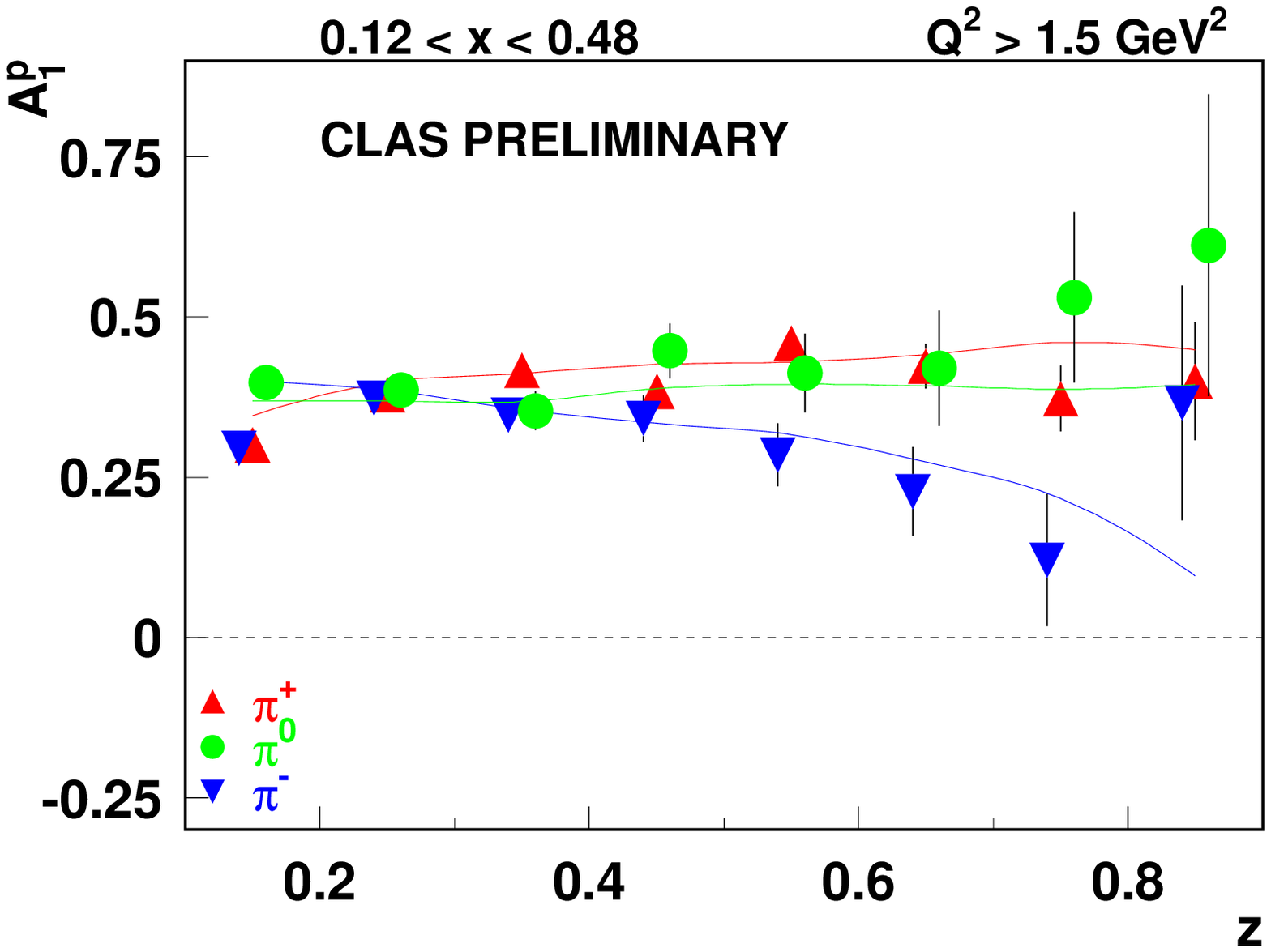}
 \end{minipage}
     \ \hspace{5mm} \hspace{0mm} \
 \begin{minipage}[b]{6.0cm}
  \includegraphics[height=.22\textheight,width=3.0in]{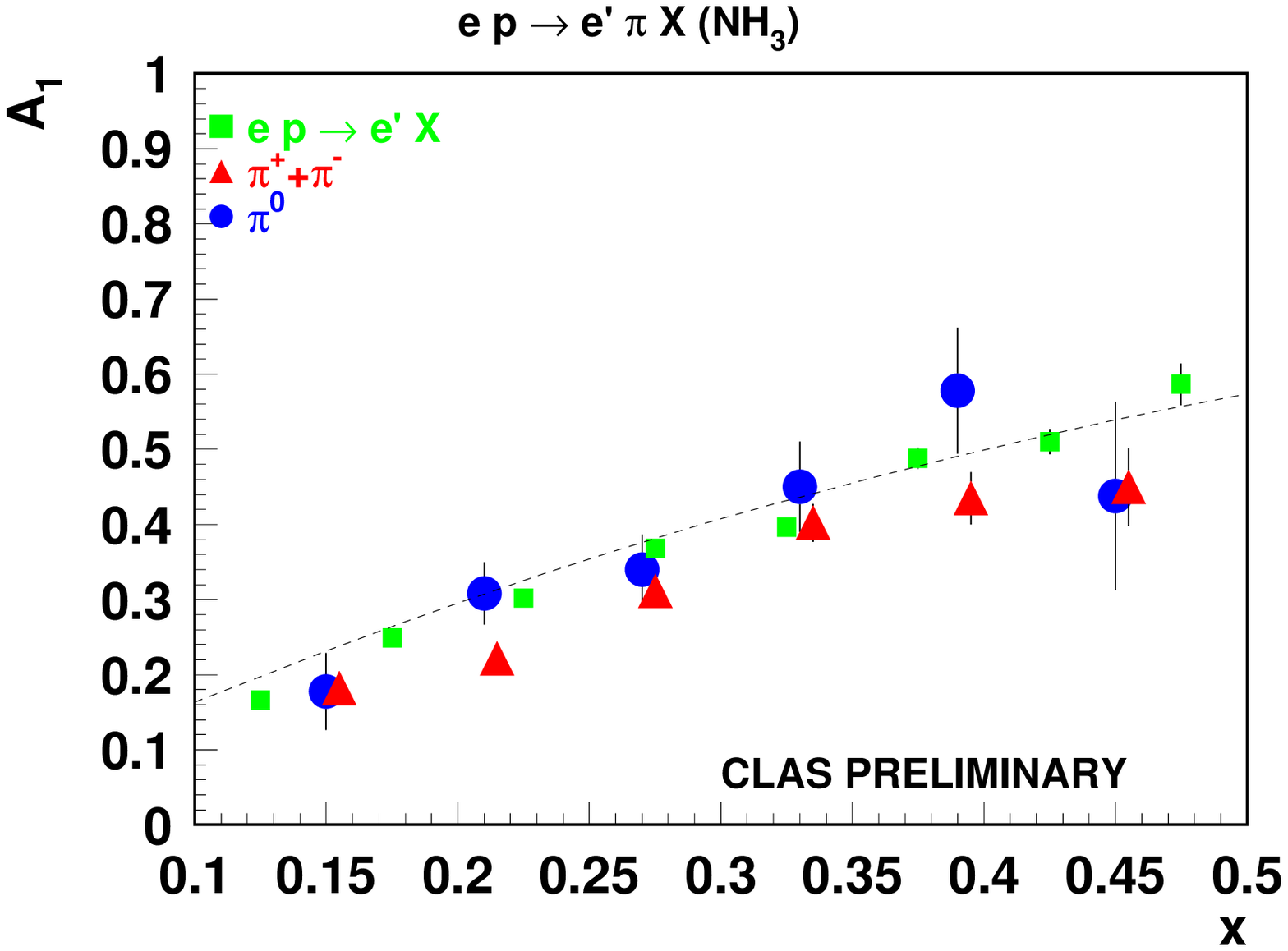}
 \end{minipage}
\caption{\small 
The double spin asymmetry for SIDIS  $\pi^+$, $\pi^-$, and  $\pi^0$  
data as a function of $z$ (left) and the comparison of inclusive $A_1^p$
with the $A_1^p$ for the SIDIS sum of  $\pi^+$ 
and  $\pi^-$ and   $A_1^p$ for SIDIS $\pi^0$ (right).}
\label{a1p3pionsxq47}
\end{figure} 
}

The validity of factorization
is crucial to the interpretation of target SSAs in terms of
TMDs, which is the main goal of this contribution.
A factorization test comes from examining the $z$-dependence
of the double spin asymmetries ($A_1^p$) for all three pion flavors, 
as shown in the left panel of Fig.~\ref{a1p3pionsxq47}.
The data cuts included $W>2$ GeV and $Q^2>1.1$ GeV$^2$ to ensure
DIS kinematics, and the average value of $x$ is approximately 0.3. 
If factorization holds, the asymmetries
should be approximately independent of $z$, broken by the
different weights given to the polarized $u$ and $d$ quarks
by the favored and unfavored fragmentation functions. We therefore
expect the largest $z$-dependence for the $\pi^-$ asymmetries.
This is indeed born out by the data, which are in excellent
agreement in both magnitude and $z$-dependence up to $z=0.7$ 
with predictions of the polarized 
Lund Monte Carlo using GRSV polarized PDFs as input. 

The dependence of the double spin asymmetry on Bjorken $x$ for 
different pion flavors
obtained from the CLAS data for the same kinematic range 
is presented in Fig. \ref{a1p3pionsxq47} (right panel).
The $\pi^0$ double spin asymmetry as well as $A_1^p$ for the sum
of charged pions are consistent with the inclusive  $A_1^p$ 
as expected in a simple partonic picture.

These studies  suggest that 
factorization works for $W>2$ GeV, $Q^2>1.1$ GeV$^2$, 
$0.15<x<0.5$, and $0.3<z<0.7$
for a 5.7 GeV electron energy.

{\vskip -3.4cm
\begin{figure}[hbtp]
\hspace{-20mm}
\includegraphics[height=.35\textheight,width=5in]{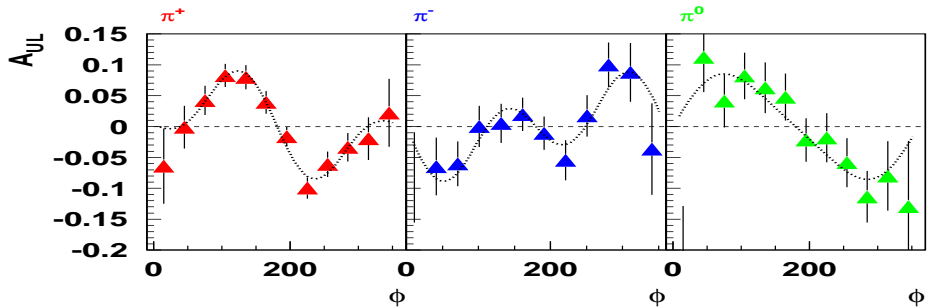}
\caption{\small 
        The target SSA as a function of azimuthal angle $\phi$ from
data at 5.7 GeV.}
\label{fig:aulsinpippi0}
\end{figure}
}

The spin-dependent moments ($\sin\phi,\sin2\phi$)
of the semi-inclusive cross section have been extracted in a fit of the 
normalized-yield asymmetry

 \begin{equation}
	\label{asy}
   A_{UL}(\phi)=\frac{1}{P_T}\frac{N^+ - N^-}{N^+ + N^-}.
 \end{equation}
\noindent  Here  $N^\pm$   
is the number of events for target polarizations
 antiparallel/parallel to the incoming beam direction  
and $P_T$ is the target polarization.

Measurements of the $\sin2\phi$ SSA 
allow the study of the Collins effect with no contamination from other
mechanisms. A recent measurement of $\sin 2\phi$ moment of $\sigma_{UL}$ by
  HERMES \cite{HER} is consistent with zero. 
A measurably large asymmetry has been predicted only at large $x$ ($x>0.2$),
 a region well-covered by JLab \cite{EFRPI0}. 

The data for $\pi^+$  (Fig. \ref{fig:aulsinpippi0}) 
show a clear $\sin\phi$  and $\sin 2\phi$ modulations
from which a $\sin\phi$ moment  of $0.058\pm0.011(stat)$ 
and $\sin 2\phi$ moment of $-0.041\pm0.011(stat)$ 
have been determined.

{\vskip -0.5cm
\begin{figure}[h]
\hspace{-20mm}
 \begin{minipage}[b]{6.5cm}
  \includegraphics[height=.22\textheight,width=4.0in]{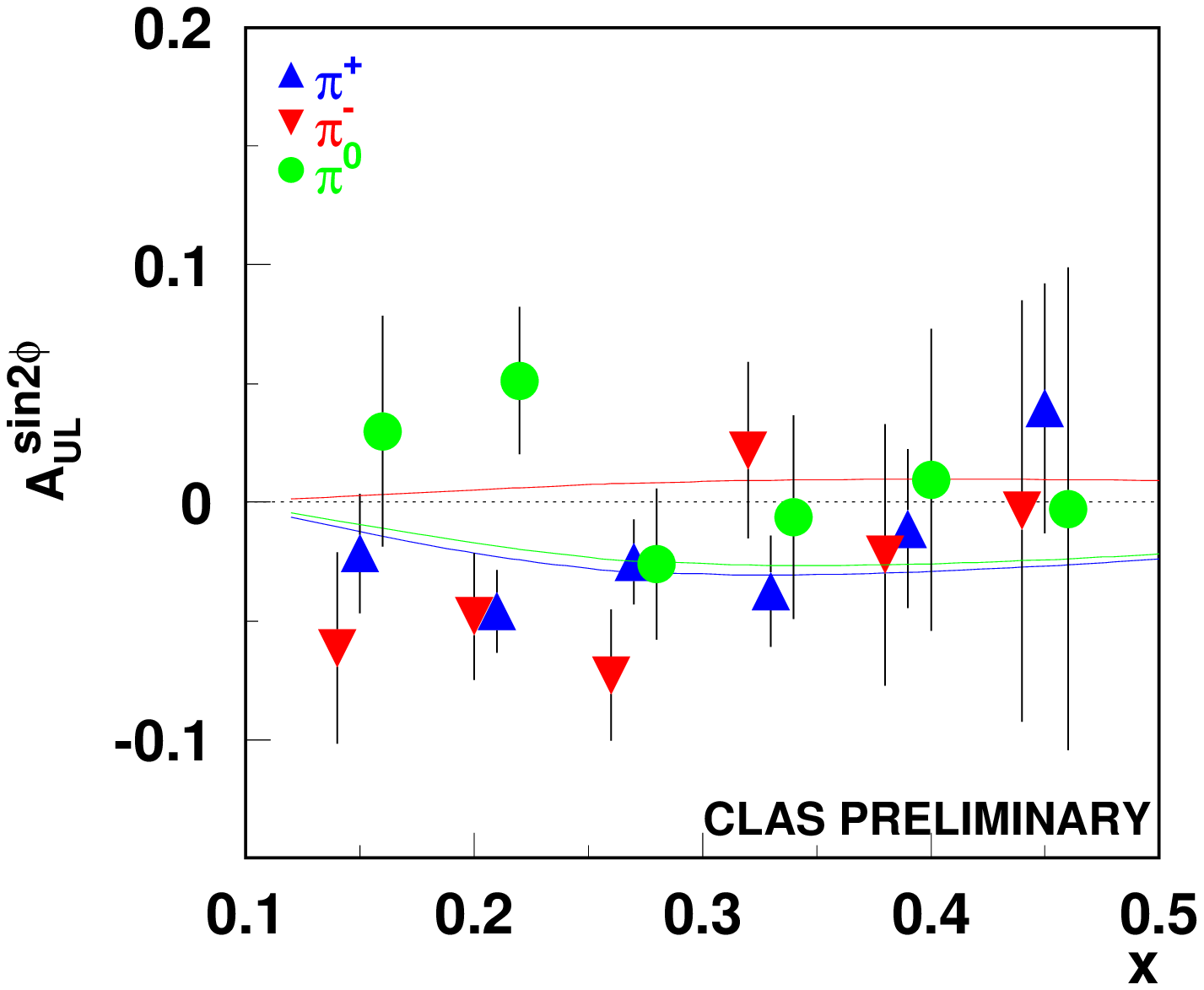}
 \end{minipage}
     \ \hspace{4mm} \hspace{0mm} \
 \begin{minipage}[b]{6.5cm}
  \includegraphics[height=.2\textheight,width=3.0in]{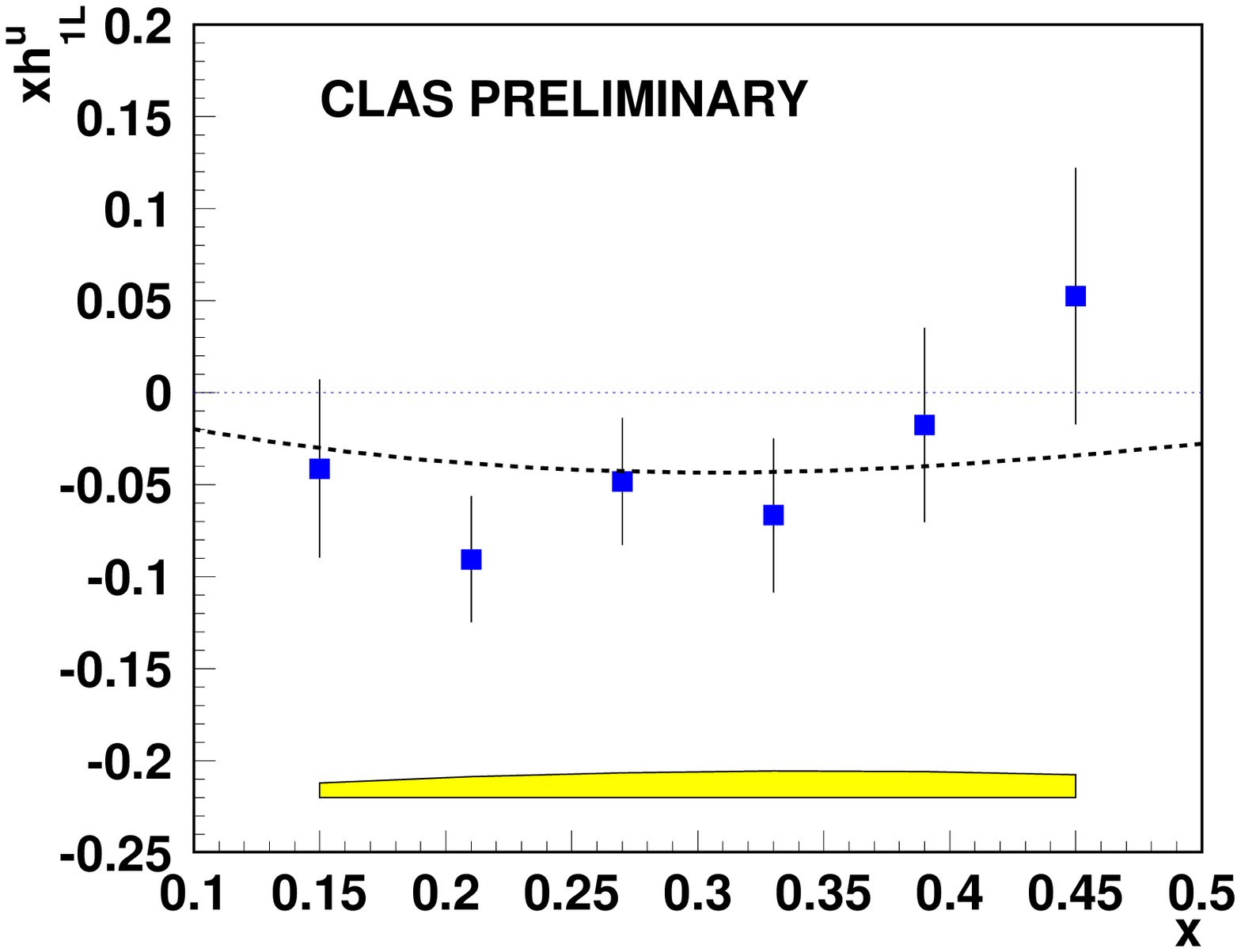}
 \end{minipage}
\caption{\small 
The leading twist SSA for the $\sin 2\phi$ moment for $\pi^+$, 
$\pi^0$, and $\pi^-$ as a function of $x$ (left plot).
The  $h_{1L}^{\perp}$ from the $\pi^+$ SSA (right). 
The contribution from the unfavored production is 
included in the systematic error band. The
variation for the ratio of unfavored to favored Collins functions is 
from -2.5 to 0.  The curves are from \cite{EFRPI0} 
using the $\chi$QSM calculations of $h_{1L}^{\perp}$.}
\label{a1pssa}
\end{figure} 
}

The $x$ dependence of the SSA
for $\pi^+$ (Fig.~\ref{a1pssa}) is consistent 
with predictions \cite{EFRPI0}. No sign of a large unfavored Collins 
fragmentation (large $\pi^-$ SSA with a corresponding $\pi^+$ SSA of opposite sign) is seen.
The $\pi^+$ SSA is dominated by the u-quarks; therefore with
some assumption about the ratio of unfavored to favored Collins fragmentation
functions, it can provide a first glimpse of the twist-2 
TMD function $h_{1L}^\perp$ (Fig. \ref{a1pssa}.) The curve is the calculation by 
Efremov et al. \cite{EFRPI0}, using $h_{1L}^\perp$ from the chiral 
quark soliton model evolved to $Q^2$=1.5 GeV$^2$.
The extraction, however, 
suffers from low statistics and has a significant systematic
error from the unknown ratio of the Collins favored and unfavored 
fragmentation functions, the unknown ratio of $h_{1L}^{\perp d}/h_{1L}^{\perp u}$,
 as well as from background from exclusive vector mesons. 

\begin{figure}[h]
  \includegraphics[height=.3\textheight,width=6in]{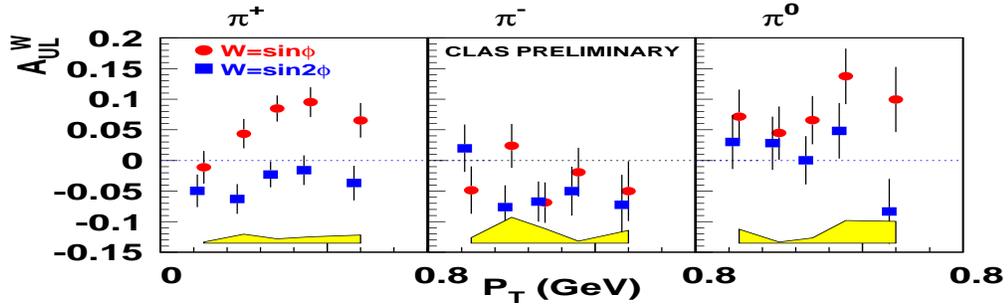}
\caption{\small 
The $A_{UL}$ SSA  dependence on $P_\perp$. The band 
represents the systematic uncertainty from fit in the measurement of the $sin2\phi$ moment.}
\label{aulx3pi}
\end{figure}

The $\sin \phi$ moment of the cross section measured with CLAS at 5.7 GeV is
in agreement with  the HERMES measurement at 27.5 GeV for a 
longitudinal target \cite{HER}, indicating  a bigger asymmetry
for the higher twist contribution compared to the leading twist $\sin2\phi$
moment.
The $P_{\perp}$ dependence of the $\sin \phi$ moment for $\pi^+$ 
(see Fig. \ref{aulx3pi}) is consistent with an increase
with increasing $P_{\perp}$, as expected for the TMD function \cite{JiMaYu}.
The $\sin \phi$ moment, for
 $\pi^-$ is also plotted in Fig. \ref{aulx3pi}, showing first evidence for
a non-zero SSA asymmetry for $\pi^-$ on a longitudinally polarized target.

The $\sin\phi$ moment of the SIDIS cross section itself can be an
important source of independent information on the Collins fragmentation 
mechanism. 
Several other contributions to the $\sin\phi$ moment 
were identified recently \cite{BoeMulPij,bac,metz}, involving different unknown distribution
and fragmentation functions. A global analysis of beam and target SSA may be required
to separate different contributions.
Drell-Yan process with  longitudinally and transversely polarized protons scattering 
on longitudinally polarized protons will also provide additional information
on these distribution functions \cite{JAF}.

In conclusion CLAS target SSAs were measured using the 5.7 GeV data
from the CLAS experiment using a polarized NH$_3$ target. The twist-2 
distribution function $h_{1L}^\perp$ was extracted for 
the first time using the $\pi^+$ target SSA.
No evidence has been found with CLAS for the
large unfavored Collins fragmentation indicated by HERMES \cite{HERtrans}.

\vskip -2.0cm

{
}
\end{document}